\documentclass[prb,aps,twocolumn,superscriptaddress]{revtex4-1}
\usepackage{graphicx}
\usepackage{amsmath}
\usepackage{amssymb}
\usepackage[english]{babel}
\begin{document}
\title{Supercurrent carried by non-equlibrium\\ quasiparticles in a multiterminal Josephson junction}

\author{M. P. Nowak}
\affiliation{AGH University of Science and Technology, Academic Centre for Materials and Nanotechnology, al. A. Mickiewicza 30, 30-059 Krakow, Poland}
\author{M. Wimmer}
\affiliation{QuTech, Delft University of Technology, P.O. Box 4056, 2600 GA Delft, The Netherlands}
\affiliation{Kavli Institute of Nanoscience, Delft University of Technology,
P.O. Box 4056, 2600 GA Delft, The Netherlands}
\author{A. R. Akhmerov}
\affiliation{Kavli Institute of Nanoscience, Delft University of Technology,
P.O. Box 4056, 2600 GA Delft, The Netherlands}
\date{\today}

\begin{abstract}
We theoretically study coherent multiple Andreev reflections in a biased three-terminal Josephson junction. We demonstrate that the direct current flowing through the junction consists of supercurrent components when the bias voltages are commensurate. This dissipationless current depends on the phase in the superconducting leads and stems from the Cooper pair transfer processes induced by non-local Andreev reflections of the quasiparticles originating from the superconducting leads. We identify supercurrent-enhanced lines in the current and conductance maps of the recent measurement [Y. Cohen, et al., PNAS {\bf 115}, 6991 (2018)] on a nanowire Josephson junction and show that the magnitude of the phase-dependent current components is proportional to the junction transparency with the power corresponding to the component order.
\end{abstract}

\maketitle
\section{Introduction}
Upon biasing of a superconductor-normal-superconductor (SNS) junction the supercurrent carried by Andreev bound states changes into quasiparticle-driven dissipative current carried in a process of multiple Andreev reflections (MAR). This phenomenon was first encountered in superconducting niobium contacts
[\onlinecite{barnes_tunneling_1969}], later in microbridges [\onlinecite{gregers-hansen_subharmonic_1973, flensberg_subharmonic_1989}] and tunnel junctions [\onlinecite{rowell_excess_1968, kleinsasser_observation_1994, van_der_post_subgap_1994}]. Klapwijk, Blonder, and Tinkham explained [\onlinecite{klapwijk_explanation_1982}] that $n$ successive Andreev reflections of electrons and holes propagating through the normal part of the junction results in charge transfer of $~(n + 1)e$, with $n/2$ Cooper pairs and a single quasiparticle between the superconductors. 
As a result the current-voltage (I-V) characteristic of the junction is imprinted with a subgap structure with features appearing for the bias voltages $V = 2\Delta/ne$ with $\Delta$ the superconducting gap. 
The original MAR theory of Ref. \onlinecite{klapwijk_explanation_1982} was soon extended allowing for description of non-transparent junctions [\onlinecite{octavio_subharmonic_1983}], coherent regime [\onlinecite{averin_ac_1995, bratus_theory_1995}]---with the prediction of appearance of DC and AC current components---, and finally the case of several current-carrying modes [\onlinecite{bardas_electron_1997}]. 
The latter allowed to estimate transmission probabilities and number of the quantized modes of atomic-thick break junctions [\onlinecite{scheer_conduction_1997, riquelme_distribution_2005}]---the pincode of the structure. Nowadays probing of MAR features became a standard technique for estimating the superconducting gap, evaluation of the number and transmission probability of conducting modes in the state-of-the-art semiconductor-superconductor hybrids that base on  nanowires [\onlinecite{gunel_supercurrent_2012, nilsson_supercurrent_2012, goffman_conduction_2017, gul_hard_2017, de_vries_spinorbit_2018}] and two-dimensional electron gas [\onlinecite{kjaergaard_transparent_2017}].

Recent progress in fabrication of hybrid nanodevices driven by the pursuit for creation topological quantum gates [\onlinecite{alicea_non-abelian_2011, heck_coulomb-assisted_2012, hyart_flux-controlled_2013}] led to creation of tunable multiterminal systems based on crossed nanowires [\onlinecite{plissard_formation_2013, fadaly_observation_2017, krizek_field_2018, vaitiekenas_selective-area-grown_2018}] or gated graphene [\onlinecite{draelos_supercurrent_2018}]. Termination of such structures by superconducting leads allows to form multiterminal Josephson junctions that serve as superconducting beam splitters that entangle Cooper pairs [\onlinecite{freyn_production_2011, jonckheere_multipair_2013}], allow to obtain Shapiro steps [\onlinecite{cuevas_voltage-induced_2007}] due to voltage-induced supercurrent and where the transconductance due to AC Josephson effect is quantized in units of $4e^2/h$ [\onlinecite{eriksson_topological_2017, deb_josephson_2018}]. Extensive work on the description of the phase dependence of Andreev bound states in multiterminal junctions [\onlinecite{virtanen_phase_2007, padurariu_closing_2015, vischi_coherent_2017}] was done demonstrating that the superconducting phases can lift Kramers degeneracy [\onlinecite{van_heck_single_2014}] or even that such junctions can be considered as an effective topological materials themselves [\onlinecite{riwar_multi-terminal_2016, strambini_-squipt_2016}].

In a Josephson junction, where a central superconducting electrode is connected with two outer superconductors solely through two separate normal regions (Josephson bi-junction), the transport between the outer S leads is possible only via Andreev reflection at the central superconductor. In this geometry the nonlocal Andreev reflections correlate four particles by the exchange of two Cooper pairs between the superconducting leads which results in a quartet supercurrent [\onlinecite{melin_partially_2016}]. Measurements on such a junction realized in a recent experiment [\onlinecite{cohen_nonlocal_2018}] found current amplification for commensurate bias voltages in line with the above theoretical prediction. The same experiment studied transport through a three-terminal structure, where all three superconducting leads are connected by a {\it common} semiconducting part [\onlinecite{cohen_non-local_2016, cohen_nonlocal_2018}], extending the previous measurement of metallic junction of the same geometry [\onlinecite{pfeffer_subgap_2014}]. The experimental maps of the current in the three-terminal nanowire-device [\onlinecite{cohen_non-local_2016}] reveal both: lines of amplified current due to MAR processes and lines analogous to those obtained in the bi-junction geometry in-line with the prediction of current spikes as a counterpart of voltage induced Shapiro steps [\onlinecite{cuevas_voltage-induced_2007}]. Despite the experimental progress the theoretical description of biased multi-terminal Josephson junctions was so far limited to incoherent [\onlinecite{houzet_multiple_2010}] and diffusive [\onlinecite{cuevas_voltage-induced_2007}] regimes or to a single bias voltage [\onlinecite{lantz_phase-dependent_2002, riwar_cross-correlations_2016, deb_josephson_2018}]  hindering the interpretation of the experimental data.

Here we develop a theory for coherent MAR in a three-terminal Josephson junction of arbitrary transparency whose leads can be biased independently. Our approach accounts for competing processes of local and non-local MAR between all the leads which allows to capture both MAR and supercurrent features in the DC response of a multi-terminal junction. We show that when the voltages are commensurate a phase-dependent current may flow without a change in voltage reminiscent of the supercurrent. This happens despite the fact that when all the bias voltages are unequal, the junction has no conserved quantities, and therefore every state in the junction has a finite lifetime and dissipates energy. We show that MAR DC current consists of a voltage-dependent, dissipative component, and series of phase-dependent, dissipationless contributions. As a result the DC current maps versus the bias voltages consist of pronounced lines of enhanced current for commensurate voltages as observed in the experiment Ref. [\onlinecite{cohen_non-local_2016, cohen_nonlocal_2018}]. We find that supercurrent components have oscillatory dependence on the superconducting phase with a period and amplitude inversely proportional to the component order.  Furthermore, we present how the visibility of the supercurrent features in the conductance maps depends on the transparency of the normal part which is applicable for analysis of the transport properties of nanoscale multiterminal devices [\onlinecite{plissard_formation_2013, fadaly_observation_2017, gazibegovic_epitaxy_2017}]

This paper is organized in the following way. The theory is given in the second section. Section III contains results of the model. Discussion of the results along with the summary is provided in sections IV and V respectively.

\section{Theory}
\subsection{Calculation of the current}
\begin{figure}[h!]
\center
\includegraphics[width = 7cm]{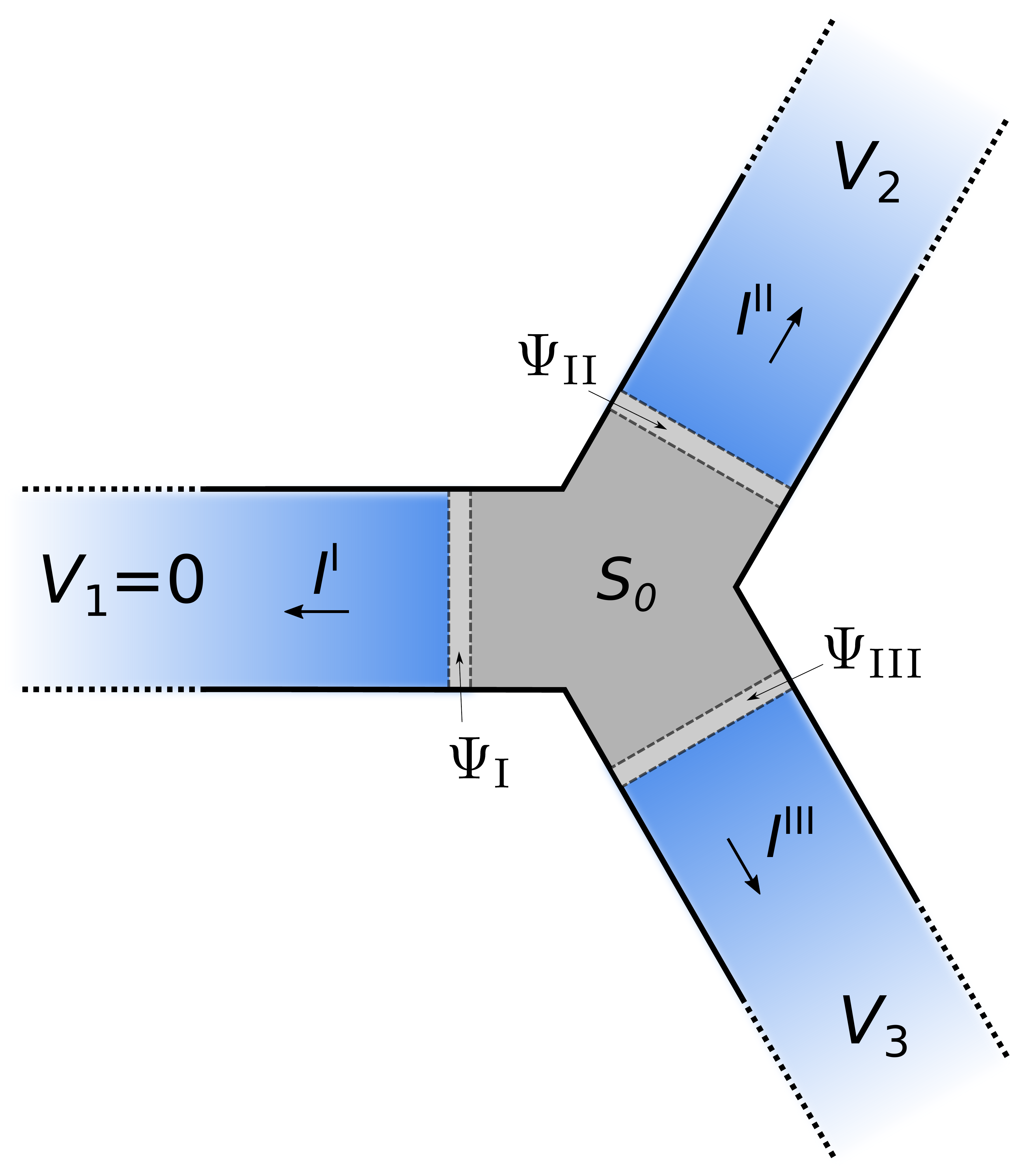}
\caption{Schematics of the considered device with three semi-infinite superconducting leads (blue) and normal scattering region (grey).}
\label{trijunction}
\end{figure}

We consider a junction that consists of three semi-infinite superconducting electrodes (S) connected by a normal region (N) depicted schematically in Fig \ref{trijunction}. We assume that the first lead is kept at voltage $V_1 = 0$, while the second and third leads are biased by $V_2$ and $V_3$ voltages respectively.

To calculate the current running through the junction we generalize the approach previously applied for a two-terminal case [\onlinecite{averin_ac_1995}] and consider quasiparticle wave-functions ($\Psi_L$) in a form of a linear combination of plane waves propagating in the normal region, adjacent to the $L$'th superconducting lead [see Fig. \ref{trijunction}],
\begin{equation}
\begin{split}
\Psi_L = \sum_{n,m} \left[ \left(
  \begin{array}{c}
    A_{n,m}^L\\
    B_{n,m}^L
  \end{array} \right) e^{ikx}+
    \left(\begin{array}{c}
    C_{n,m}^L\\
    D_{n,m}^L
  \end{array} \right) e^{-ikx}
  \right]\times \\e^{-i\left[E+neV_2+meV_3\right]t/\hbar}.
\end{split}
\end{equation}
The time dependence accounts for the voltage applied to the superconducting leads, $A_{n,m}^L$, $C_{n,m}^L$ ($B_{n,m}^L$, $D_{n,m}^L$) correspond to electron (hole) amplitudes and $x$ points in the direction opposite to the scattering region for all three leads.

The scattering properties of the normal part of the junction are contained within the scattering matrix $S_0$ that is used to setup the matching conditions for the wave-functions $\Psi_L$. For the electron part we have:
\begin{equation}
\left(\begin{array}{c}
    A_{n,m}^\text{I}\\
    A_{n+1,m}^{\text{II}}\\
    A_{n,m+1}^{\text{III}}\\
  \end{array}\right) = S_{0} \left(
  \begin{array}{c}
    C_{n,m}^\text{I}\\
    C_{n+1,m}^\text{II}\\
    C_{n,m+1}^\text{III}
  \end{array} \right),
\label{sle1}
\end{equation}
and for the hole part:
\begin{equation}
\left(\begin{array}{c}
    D_{n,m}^\text{I}\\
    D_{n-1,m}^\text{II}\\
    D_{n,m-1}^\text{III}\\
  \end{array}\right) = S_{0} ^*\left(
  \begin{array}{c}
    B_{n,m}^\text{I}\\
    B_{n-1,m}^\text{II}\\
    B_{n,m-1}^\text{III}
  \end{array} \right),
\label{slh1}
\end{equation}
with the shifts of the indices that account for the particles gaining and loosing energy due to the bias voltages.

At each SN interface we account for Andreev reflection and acquisition of the phase present at the superconducting lead, hence:
\begin{equation}
\left(\begin{array}{c}
    C_{n,m}^L\\
    B_{n,m}^L
  \end{array}\right)
  = \sigma_{L}\left(\begin{array}{c c}
    a_{n,m} & 0\\
    0 & a_{n,m}
  \end{array} \right)
  \left(
  \begin{array}{c}
    D_{n,m}^L\\
    A_{n,m}^L
  \end{array} \right),
\label{AA}
\end{equation}
where the Andreev-reflection amplitudes are given by $a_{n,m}\equiv a(E+neV_2+meV_3)$, with,
\begin{equation}
a(E) = \frac{1}{\Delta}\left\{
  \begin{array}{l l}
    E - \mathrm{sgn}(E)\sqrt{E^2-\Delta^2} & \quad |E|>\Delta\\
    E - i \sqrt{\Delta^2-E^2} & \quad |E| \leq \Delta
  \end{array} \right.,
\end{equation}
and where
\begin{equation}
\sigma_L = \left(
  \begin{array}{c c}
    e^{-i\phi_L} & 0\\
    0 & e^{i\phi_L}
  \end{array} \right),
\end{equation}
accounts for the phase shift at the $L$'th SN interface.

An electron/hole excitation in the normal part of the junction is created by incoming quasiparticles from the nearby superconductors. We assume that there is no Fermi wavelength difference between the normal and superconducting parts and that the chemical potential is much higher than the energy gap. In the superconductor for the energies exceeding the superconducting gap there are two modes propagating towards the normal region [\onlinecite{datta_scattering_1996}] with the wave-functions:
\begin{equation}
\Psi_{inc}^{qe} = \left( \begin{array}{c}
    u\\
    v
  \end{array} \right) e^{-ikx},\;
  \Psi_{inc}^{qh} = \left( \begin{array}{c}
    v\\
    u
  \end{array} \right) e^{ikx},
\end{equation}
where $\Psi_{inc}^{qe}$ ($\Psi_{inc}^{qh}$) corresponds to the quasiparticle that has electron-like (hole-like) character for $|E| \gg \Delta$. The corresponding amplitudes $u,v$ are $\left[\left( 1+\sqrt{1-(\Delta/E)^2}\right)/2\right]^{1/2}$ $\left[\left( 1-\sqrt{1-(\Delta/E)^2}\right)/2\right]^{1/2}$ respectively. 

Assuming that the incoming quasiparticle has electron-like character we write the wave-functions at each side of the SN interface. At the superconducting part we have:
\begin{equation}
\Psi_S(x) = \left( \begin{array}{c}
    u\\
    v
  \end{array} \right) e^{-ikx}+
  a \left( \begin{array}{c}
    u\\
    v
  \end{array} \right) e^{ikx}+
  b \left( \begin{array}{c}
    v\\
    u
  \end{array} \right) e^{-ikx},
\end{equation}
where $a,b$ stand for the amplitudes of the reflected quasiparticles. On the normal side we have:
\begin{equation}
\Psi_N(x)=
  c \left( \begin{array}{c}
    1\\
    0
  \end{array} \right) e^{-ikx}+
  d \left( \begin{array}{c}
    0\\
    1
  \end{array} \right) e^{ikx}.
\end{equation}

Matching the wave-functions and their derivatives at the interface between the materials ($x=0$) one finds $a=d=0$, $c=\frac{u^2-v^2}{u}$. Taking into account the superconducting density of states ($1/\sqrt{u^2-v^2}$) the electron-like quasiparticle with energy $E$, creates an electron in the normal part with the wave-function amplitude $J = \frac{\sqrt{u^2-v^2}}{u} \sqrt{F_D(E)}$ with ${F_D(E)}$ the Fermi distribution that determines the filling of the electron band. Following the same procedure one finds that hole-like quasiparticle creates a hole in the normal part with the same amplitude.

The above relations allow us to write Eq. \ref{AA} including the source terms:
\begin{equation}
\begin{split}
\left(\begin{array}{c}
    C_{n,m}^{L}\\
    B_{n,m}^{L}
  \end{array}\right)
  =& \sigma_{L}\left(\begin{array}{c c}
    a_{n,m} & 0\\
    0 & a_{n,m}
  \end{array} \right)
  \left(
  \begin{array}{c}
    D_{n,m}^{L}\\
    A_{n,m}^{L}
  \end{array} \right)\\&+
  \left( \begin{array}{c}
    J(E+eV_L)\\
    0
  \end{array} \right)\frac{1}{\sqrt{2}}\delta_{p,e}\delta_{s,2}\kappa_{L}^+\\&+
  \left(\begin{array}{c}
    0\\
    J(E-eV_L)
  \end{array} \right)\frac{1}{\sqrt{2}}\delta_{p,h}\delta_{s,2}\kappa_{L}^-,
\end{split}
\label{AA_final}
\end{equation}
where $p$ controls the incoming quasiparticle type, $s$ the position of the source term, and 
$\kappa_1^\pm = \delta_{n,0}\delta_{m,0}$, $\kappa_2^\pm = \delta_{n,\pm1}\delta_{m,0}$, $\kappa_3^
\pm = \delta_{n,0}\delta_{m,\pm1}$ account for the shifts in the chemical potential introduced by the bias voltage. 

We calculate the electric current $I$ in the $L$'th lead from the probability current taking into account one-dimensional density of states:
\begin{equation}
I^{L}=\sum_{\imath,j}^{I_\text{max}} I^{L}_{\imath,j} e^{(\imath V_2+jV_3)eit/\hbar},
\label{total_current_formula}
\end{equation}
where,
\begin{equation}
\begin{split}
I_{\imath,j}^{L}=\frac{e}{\hbar\pi}\sum_{s=1,2,3}\sum_{p=e,h}\int_{-\infty}^{\infty}dE\\\sum_{n,m}^{N_\text{max}}(\mathbf{U}^{L*}_{\imath+n,j+m}\mathbf{U}^L_{n,m}
-\mathbf{V}^{L*}_{\imath+n,j+m}\mathbf{V}^L_{n,m}),
\end{split}
\label{CII}
\end{equation}
are Fourier components of the current. $\mathbf{U}^L_{n,m}=\left(A_{n,m}^L,B_{n,m}^L\right)^T$ and $\mathbf{V}^L_{n,m}=\left(C_{n,m}^L,D_{n,m}^L\right)^T$ stand for vectors that consist of electron and hole amplitudes of wave-functions that carry positive and negative current, respectively. 

The integral is evaluated numerically, where at each point of the integration we solve system of equations built with Eqs. (\ref{sle1},\ref{slh1},\ref{AA_final}). We assume zero temperature and take $I_\text{max}={N_\text{max}}=8$ for the calculations when both voltages are varied which sets the limit to the convergence to voltages larger than $|eV| > \Delta/8$. Treatment of the regime with $V_2, V_3$ close to zero---with the current below the critical current---is beyond the scope of the present work. The code used for the calculations is available in Ref. [\onlinecite{noauthor_code_nodate}].

\subsection{Scattering matrix}
We assume that the central part of the junction is a symmetric beam splitter [grey region in Fig. \ref{trijunction}] and that the normal part of the junction is shorter than the superconducting coherence length. In other words it has an energy-independent scattering matrix:
\begin{equation}
S_{0}=\left(\begin{array}{ccc}
    \alpha & \beta & \beta \\
    \beta & \alpha & \beta \\
    \beta & \beta & \alpha
  \end{array}\right),
\label{smat}
\end{equation}
with $\alpha= -e^{i a}/\left( 2e^{i a}-e^{-i a}\right)$ and $\beta = 2i\sin(a)/\left( 2e^{ia}-e^{-ia}\right)$ [\onlinecite{itoh_scattering_1995}], where $a$ controls the transparency of the splitter with the transmission probability $D = 2|\beta|^2$.

\section{Results}
\subsection{Current versus the bias voltages}
\begin{figure}[t!]
\center
\includegraphics[width = 8cm]{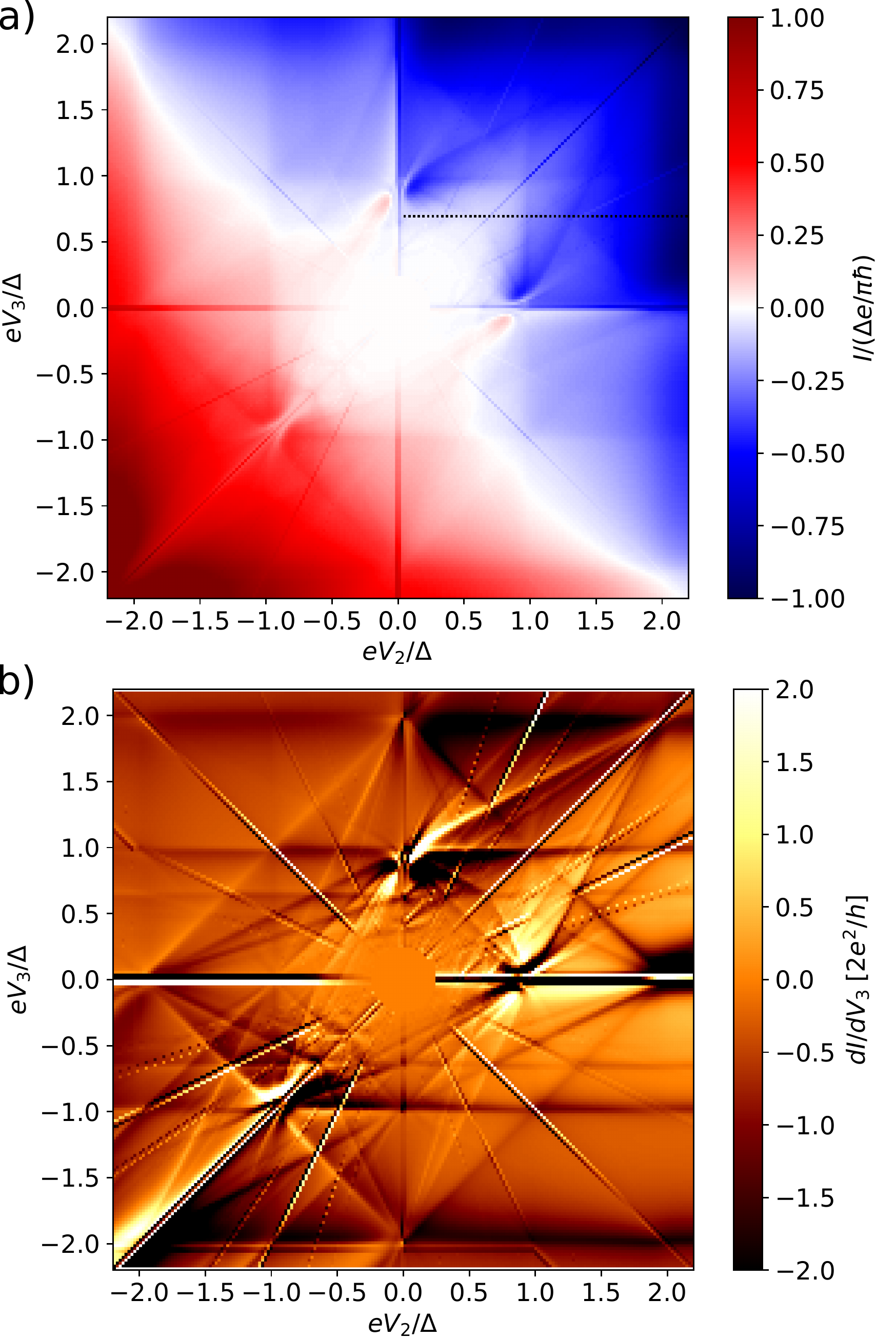}
\caption{a) DC current in the first lead for $D = 0.5$. b) Differential conductance obtained from the map of (a).}
\label{I_G_D05}
\end{figure}

Let us start by inspecting the DC response of the junction due to biasing of the second and third leads. Figure \ref{I_G_D05} (a) shows the current in the first lead while (b) presents the differential conductance obtained from the current map. In the maps we include components of the current that fulfill $\imath eV_2+jeV_3 < 5\cdot10^{-3}$ meV such the current features at commensurate voltages have comparable width to the ones observed experimentally [\onlinecite{cohen_non-local_2016, cohen_nonlocal_2018}]. We find two types of features in the conductance map: the centrifugal lines at commensurate voltages ($\imath V_2 + jV_3 = 0$) that are distinctively sharper than other subharmonic features that appear at ($\imath V_2 + jV_3 = 2\Delta/e$) due to non-local MAR. Accordingly, we observe pronounced lines of altered current at commensurate voltages $V_2 = V_3$ and $V_2 = -V_3$ in the map of Fig. \ref{I_G_D05} (a). Note, that sole time-reversal symmetry present in the considered system is not enough to guarantee the symmetry of the current with respect to the change of the sign of the voltages as application of this symmetry leads to different electron occupation in the leads.

\begin{figure}[h!]
\center
\includegraphics[width = 7.5cm]{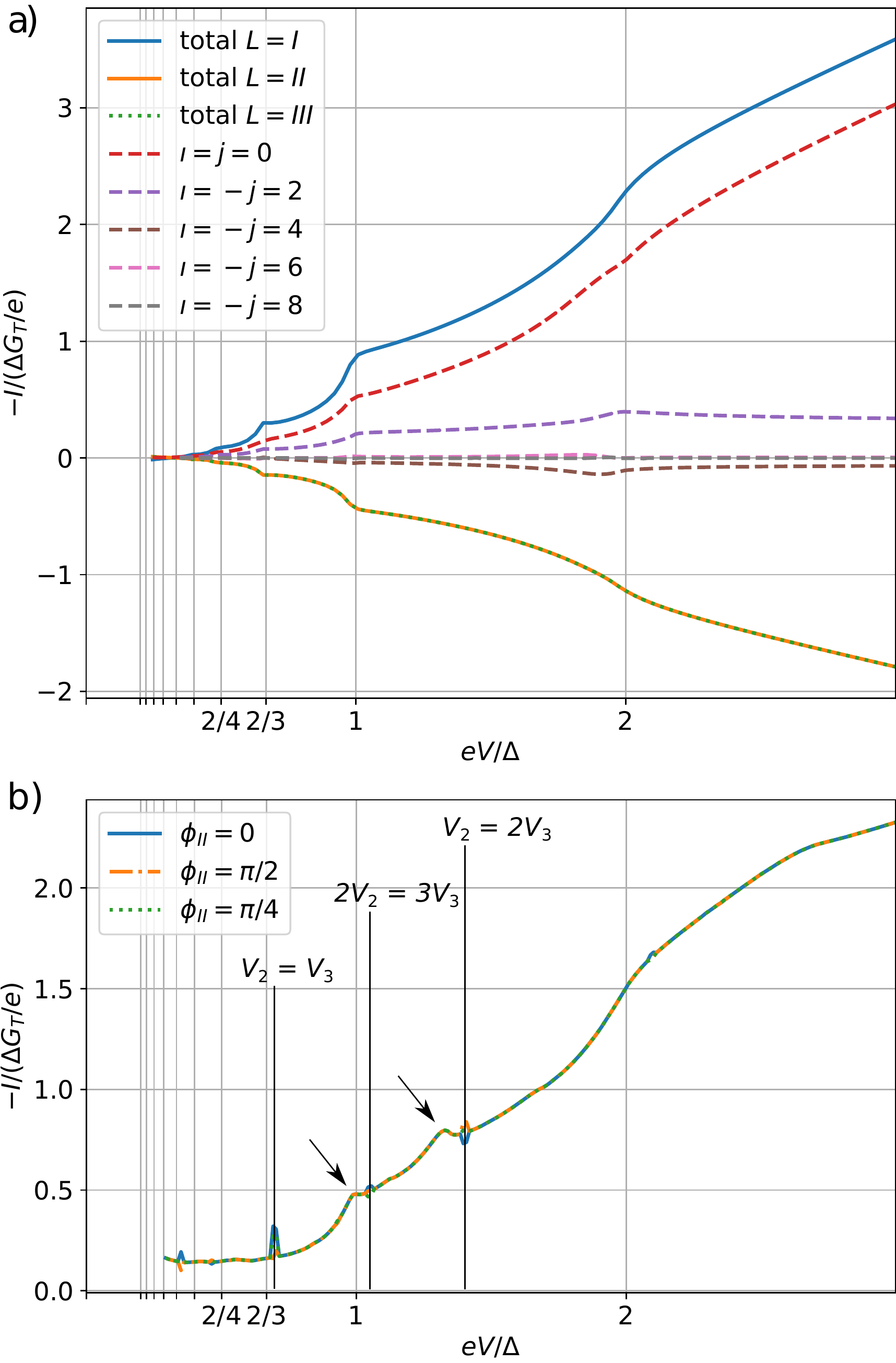}
\caption{DC current (solid and dotted curves) in the leads versus the bias energy $eV$ with $V_2 = V_3 = V$ and $D = 0.5$. Dashed curves present DC components $I_{\imath,j}^I$ of the current in the first lead. $G_T = e^2D/\pi\hbar$.  b) Cross-section of the current map of Fig. \ref{I_G_D05} (a) for $eV_3 = 0.7 \Delta$.}
\label{current_crossections}
\end{figure}

We plot the current running through the first lead in the regime $V_2=V_3=V$  with the blue curve in Fig. \ref{current_crossections} (a). In this regime the three-terminal junction reduces to a two-terminal one and the current reproduces exactly the MAR response obtained in the work of Ref. \onlinecite{averin_ac_1995}. Moreover, due to symmetry of the second and the third lead we find exactly the same current in both equally biased leads depicted with the green crosses overlapping the orange curve. The magnitude of that current is half of the current running through the first lead and hence current conservation law $\sum_{L} I^L=0$ is fulfilled. Note however that when the voltages fulfill $\imath V_2+jV_3=0$ the current consists of multiple DC components, obtained for $\imath = -j$ in Eq. (\ref{total_current_formula}). We plot the individual current components in Fig. \ref{current_crossections} (a) with dashed curves and observe that the increase of the component order $|\imath|,|j|$ results in a decrease of the current amplitude. Already the sixth-order component $\imath=-j=6$ is nearly zero. The components of odd order are zero, as the particles outgoing from and reaching the leads are shifted in energy by even multiples of the bias voltages times $e$. 

In Fig. \ref{current_crossections} (b) we present cross-section of the map of Fig. \ref{I_G_D05} (a) obtained for $eV_3 = 0.7 \Delta$. The sharp features appearing at commensurate voltages are marked by black vertical lines while the non-local MAR features at ($\imath V_2 + jV_3 = 2\Delta/e$) are marked with arrows. Upon applying of the phase difference at the biased lead the current for the commensurate voltages changes---as explicitly seen in the peak heights dependence for different values of $\phi_{\text{II}}$ in Fig. \ref{current_crossections} (b). In the following section we will discuss the origin of this phenomenon.

\subsection{Phase dependence}
\begin{figure}[h!]
\center
\includegraphics[width = 6.5cm]{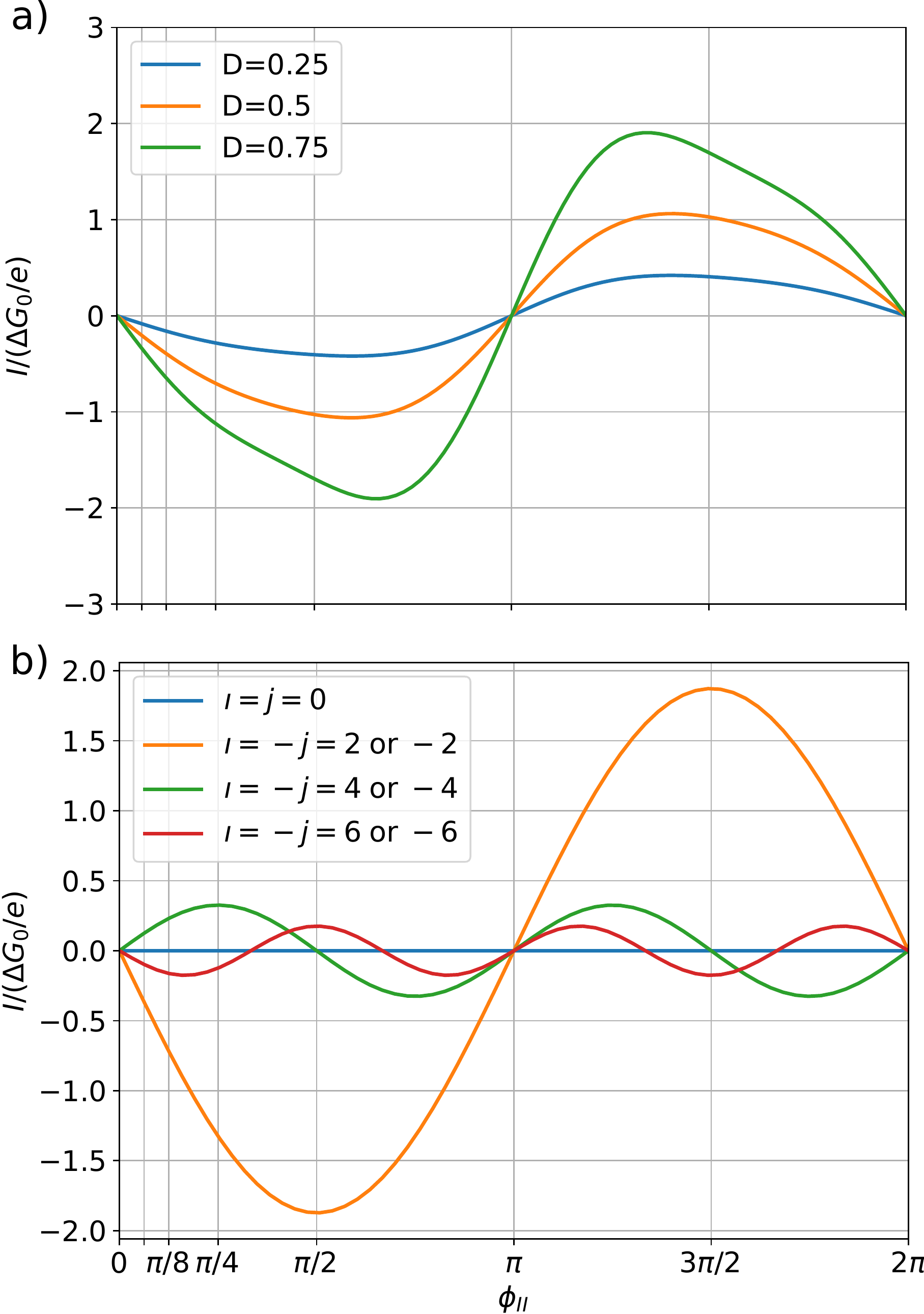}
\caption{(a) DC current running between the second and the third lead $I^\text{II}-I^\text{III}$ for three values of the central region transparency $D$ for $V_2=V_3=e\Delta/2$ versus the phase on the second lead. (b) Components  of the current for $D=0.75$. }
\label{phase_dependencies_p}
\end{figure}

Let us now inspect the phase dependence of the current carried between the second and third lead depicted in Fig. \ref{phase_dependencies_p} (a). We set $\phi_\text{I}=\phi_{\text{III}}=0$ and vary $\phi_{\text{II}}$. In agreement with the equal currents in the second and third leads of Fig. \ref{current_crossections} (a) we see zero current for $\phi_\text{II}=0$. Upon introduction of the phase difference a nonzero current arises and its phase dependence resembles the supercurrent phase-relation of a diffusive Josephson junction [\onlinecite{beenakker_universal_1991}] for all the transparencies $D$ considered here. Focusing on $D=0.75$ we decompose the current into the DC components. Remarkably, we see that the component $\imath=j=0$ is zero [the blue curve in Fig. \ref{phase_dependencies_p} (b)] as the voltage difference between the second and third lead is also zero and as the modification of the amplitudes by the superconducting phase does not contribute to the calculated current for $\imath=j=0$. The subsequent components of $\imath$'th order oscillate with the period inversely proportional to the component order with $\sim\sin(|\imath/2|\phi_\text{II})$. 

\begin{figure}[h!]
\center
\includegraphics[width = 6.5cm]{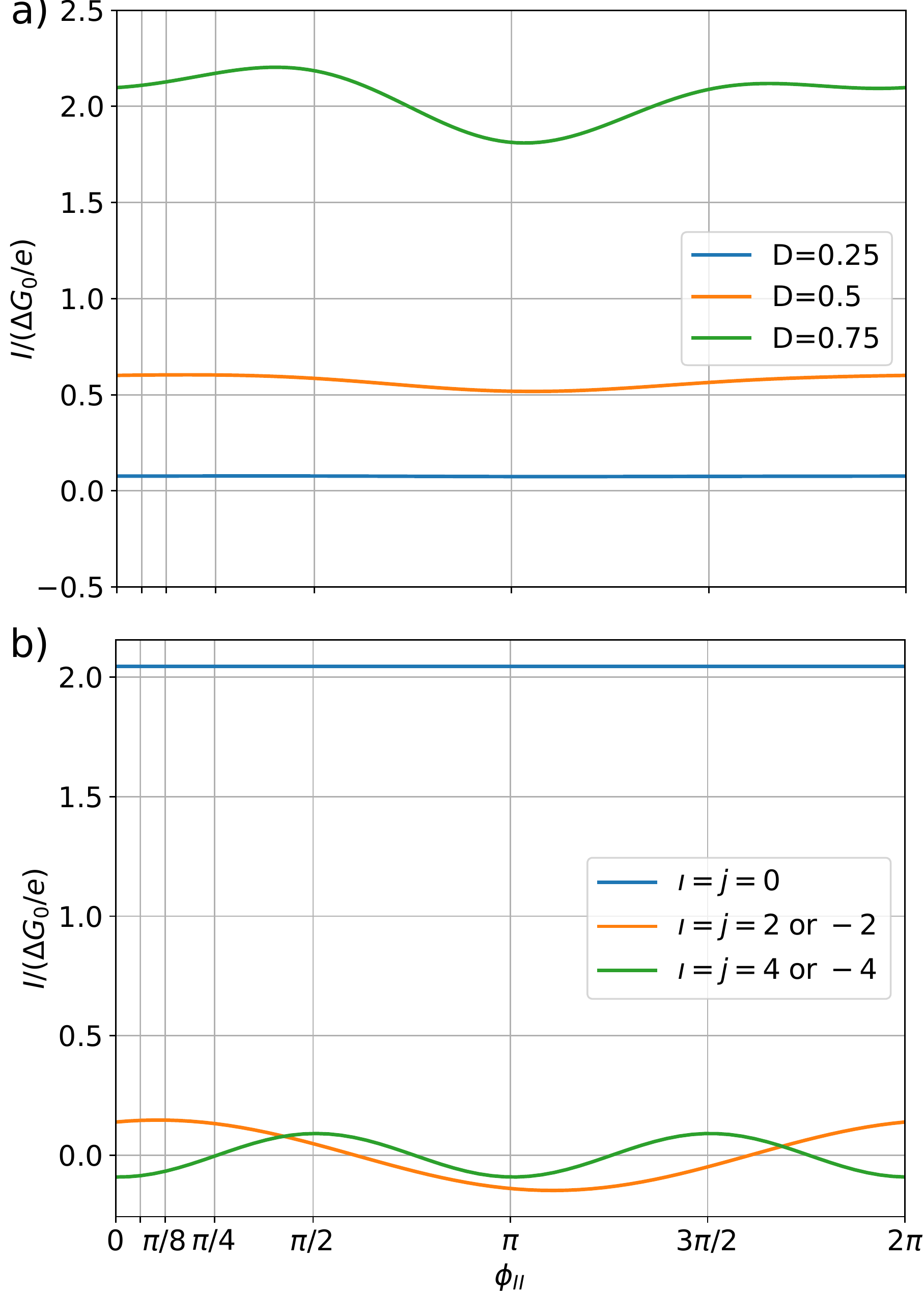}
\caption{Same as Fig. \ref{phase_dependencies_p} but for $V_2=-V_3=e\Delta/2$.}
\label{phase_dependencies_m}
\end{figure}

Similar analysis performed for another commensurate voltage configuration, i.e. $V_2=-V_3= \Delta/2e$ also shows that the current oscillation amplitude decreases when the transparency is reduced [see Fig. \ref{phase_dependencies_m} (a)]. By decomposing the current into the separate DC components we see that again the zeroth order component is phase independent and it corresponds to the ohmic current that is proportional to the voltage difference between the biased leads. The higher order components are phase-dependent and oscillate around zero with a small amplitude following the same rule of the decrease of period as the currents of Fig. \ref{phase_dependencies_p} (b). Those components then constitute the supercurrent that is carried between commensurately biased leads for commensurate voltages. The small phase offsets of the higher order current components are the result of phase shifts introduced by complex transmission and reflection amplitudes in the scattering matrix when the scattering involves non-local processes.

\subsection{Quasiparticle driven supercurrent versus the scattering paths}

\begin{figure}[h!]
\center
\includegraphics[width = 7.5cm]{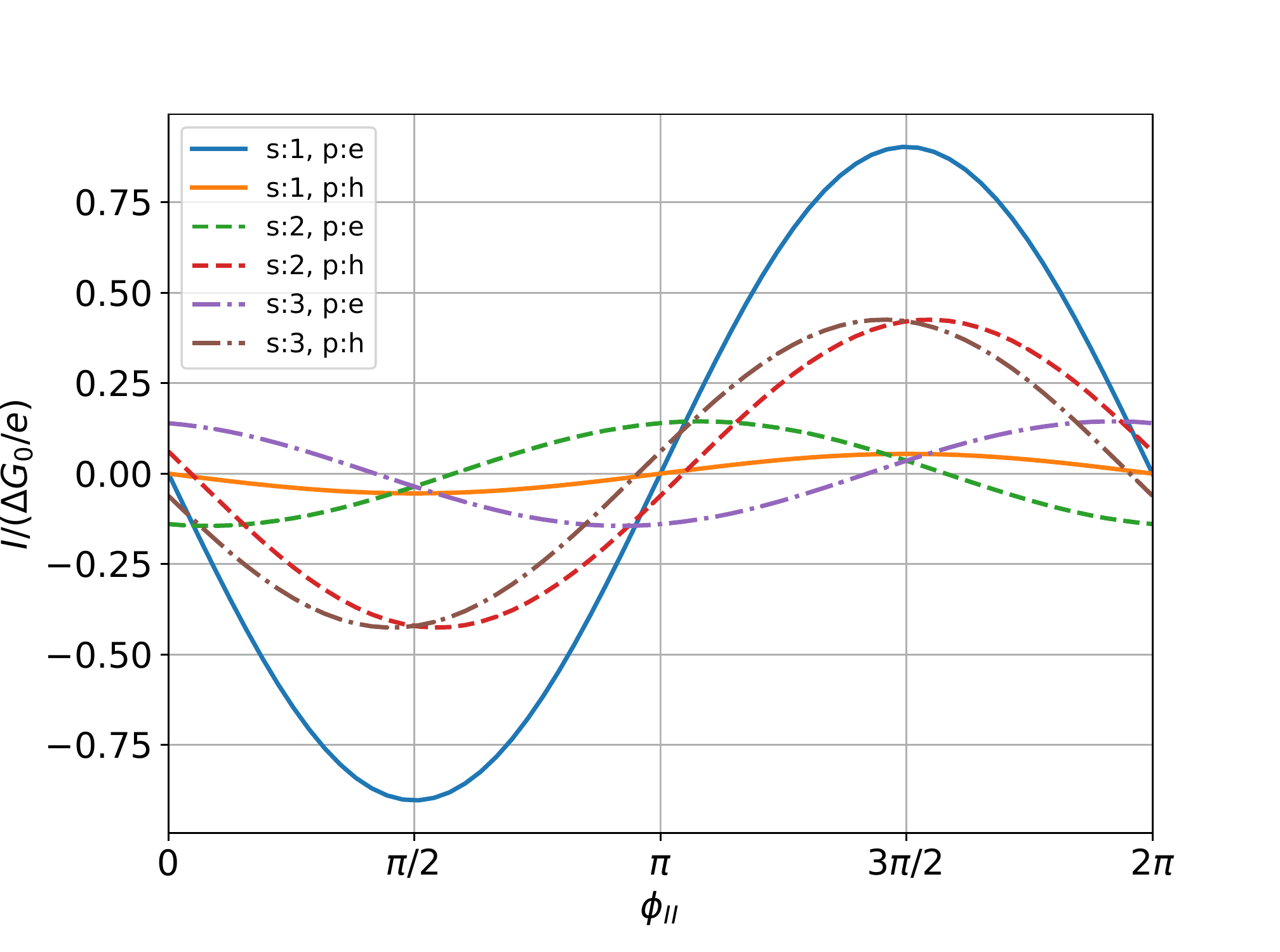}
\caption{$I^\text{II}-I^\text{III}$ DC current component $\imath=-j=2$ calculated for different source terms ($s$ -- the source position, $p$ -- quasiparticle type).}
\label{phase_dependencies_components}
\end{figure}

To understand the origin of the supercurrent let us inspect the quasiparticle scattering processes that stand behind one of the higher order components of the current running between second and the third lead, namely $\imath=-j=2$ for $V_2 = V_3 = \Delta/2e$. In the first step we isolate the source terms that that initiate the majority of the current flowing between the leads. In Fig. \ref{phase_dependencies_components} we display the current broken down into contribution from all possible source terms: $s$ corresponds to the lead where we include the source and $p$ to the quasiparticle type---see Eq. (\ref{CII}). We see that actually half of the current between equally biased leads results from an electron-like quasiparticle incoming from the {\it first} lead. Another source term that induces a considerable current correspond to hole-like quasiparticles injected from the second and third lead, which are Andreev reflected at the first lead forming an electron propagating towards the scattering region as in the case of the major component. As a result in such symmetric bias configuration the current component due to source term $(s:1,p:e)$ is equal to the sum of components due to $(s:2,p:h)$ and $(s:3,p:h)$. Similarly the sum of the current components due to source term $(s:2, p:e)$ and $(s:3, p:e)$ correspond to the current component due to $(s:1,p:h)$. 

\begin{figure}[h!]
\center
\includegraphics[width = 8.2cm]{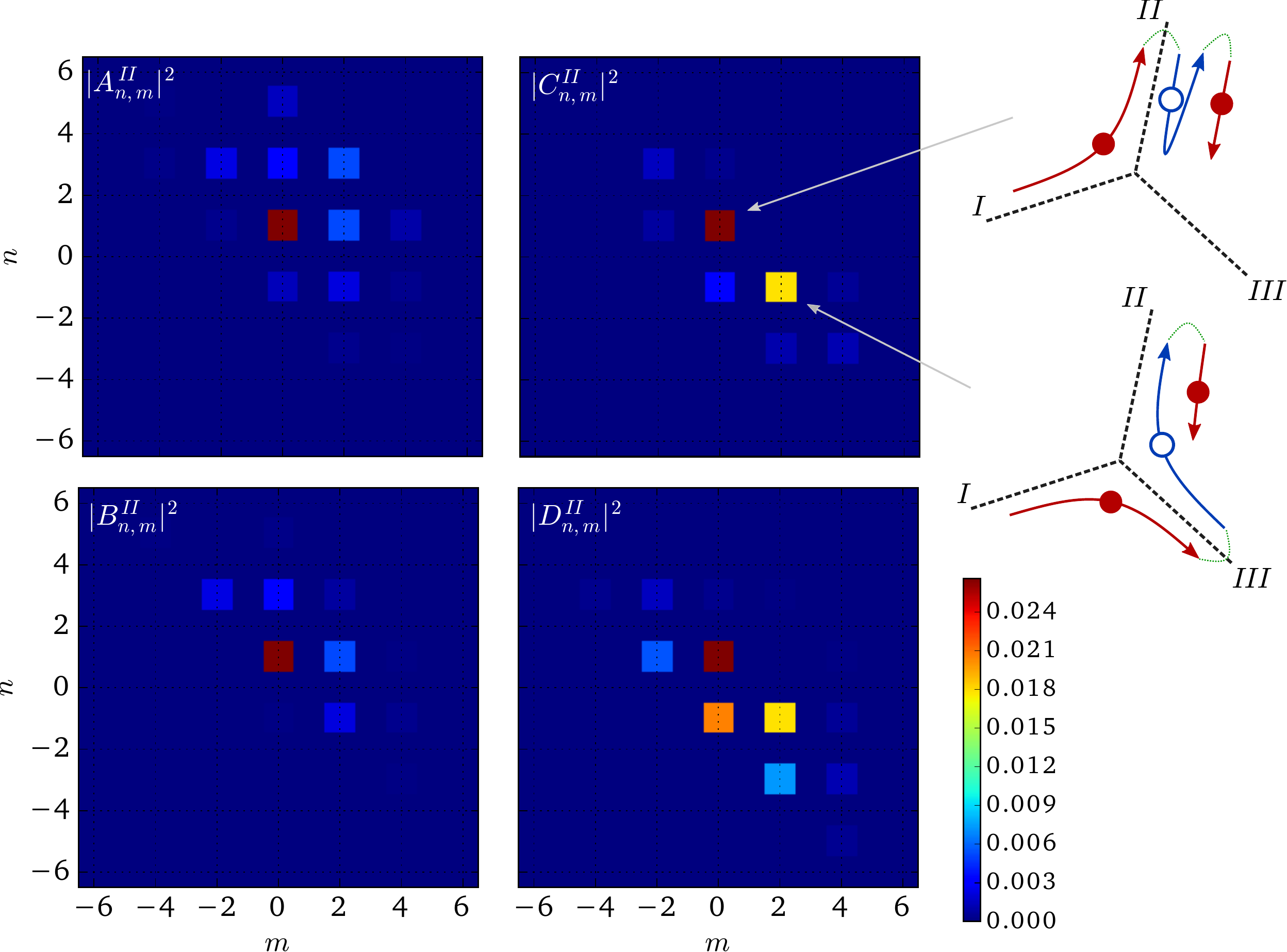}
\caption{Probabilities of the wave-function components adjacent to the second lead for electron-like quasiparticle as the source term at energy $-1.0001 \Delta$ injected from the first lead. Transparency is $D = 0.75$ and $eV_2=eV_3=\Delta/2$. The insets on the right-hand-side we show two scattering processes contributing to the second DC component (quasiparticle induced supercurrent) that result from an electron (red filled circle) propagating towards the scattering region.}
\label{amplitudes_together}
\end{figure}

Next, let us trace the scattering events that happen after the electron-like quasiparticle incomes from the first lead. We set the injection energy just below the gap, i.e. multiply the integrand in Eq. (\ref{CII}) by Dirac delta located at $E = -1.0001\Delta$. In Fig. \ref{amplitudes_together} we plot the squared absolute values of amplitudes $A^\text{II}_{n,m},B^\text{II}_{n,m},C^\text{II}_{n,m},D^\text{II}_{n,m}$ of the wave-functions adjacent the second superconducting lead. This way we are able to quantify the probability of finding electron and hole components of the wave-functions with a given propagation direction at certain energies. We observe that overall the highest probabilities correspond to the wave-functions of the electron propagating towards the scattering region [top right panel on Fig. \ref{amplitudes_together}]---with the probabilities $|C^\text{II}_{n,m}|^2$ -- and to its time reversed partner---a hole propagating in the opposite direction [bottom right panel on Fig. \ref{amplitudes_together}]---with probabilities $|D^\text{II}_{n,m}|^2$. Focusing on the electron part we see that the most important contributions to the current stem from wave-function amplitudes with indexes: $n=1,m=0$ and $n=-1,m=2$, which both give the wave-function shifted in energy by $eV_2=eV_3$. The $C_{n=1,m=0}^\text{II}$ coefficient is populated in a process where an electron coming from the first lead is Andreev reflected in the second lead, scatters back to this lead and finally is Andreev reflected again---see the top inset to Fig. \ref{amplitudes_together}. This process is insensitive to the phase on the second lead and we approximate its probability as $D(1-D)$. The finite $C_{n=-1,m=2}^\text{II}$ amplitude  stems from a process where the electron enters the scattering region from the first lead and propagates to the third lead. Due to nonlocal Andreev reflection, the retro-reflected hole propagates to the second lead where it is Andreev reflected again into an electron with the energy shifted by $2eV_3-eV_2=\Delta/2$---see bottom inset to Fig. \ref{amplitudes_together}. As a result we can estimate the corresponding probability as $\sim D^2$. The process that populates the $C_{n=-3,m=4}^\text{II}$ amplitude can be drawn analogically, only now the hole visits the second lead two times and by that gains twice the phase shift.

According to Eq. (\ref{CII}) and Eq. (\ref{total_current_formula}) the main DC component of the current $\imath=j=0$ is obtained from the sum of the {\it probabilities} set by the wave-function amplitudes and hence it cannot be directly dependent on phase of the amplitudes. On the other hand, the higher order DC current components are obtained as a sum of the products of wave-functions {\it amplitudes} with indexes $n$ and $m$ shifted by $\imath$ and $j$ respectively, such $\imath V_2+jV_3=0$ and by that they are phase-dependent. The higher order components result from the scattering processes with increased number of Andreev reflection at the phase-biased lead, resulting in a decrease of period of oscillation in phase as seen in Fig. \ref{phase_dependencies_p} (b). This means that the supercurrent (i.e. the higher order components of the DC current) originate from quasiparticles entering from the superconducting leads but also that it involves non-local Andreev reflection process of non-equlibrium electron/holes at the biased superconductor leads.


\subsection{Transparency dependence}
\begin{figure}[h!]
\center
\includegraphics[width = 8cm]{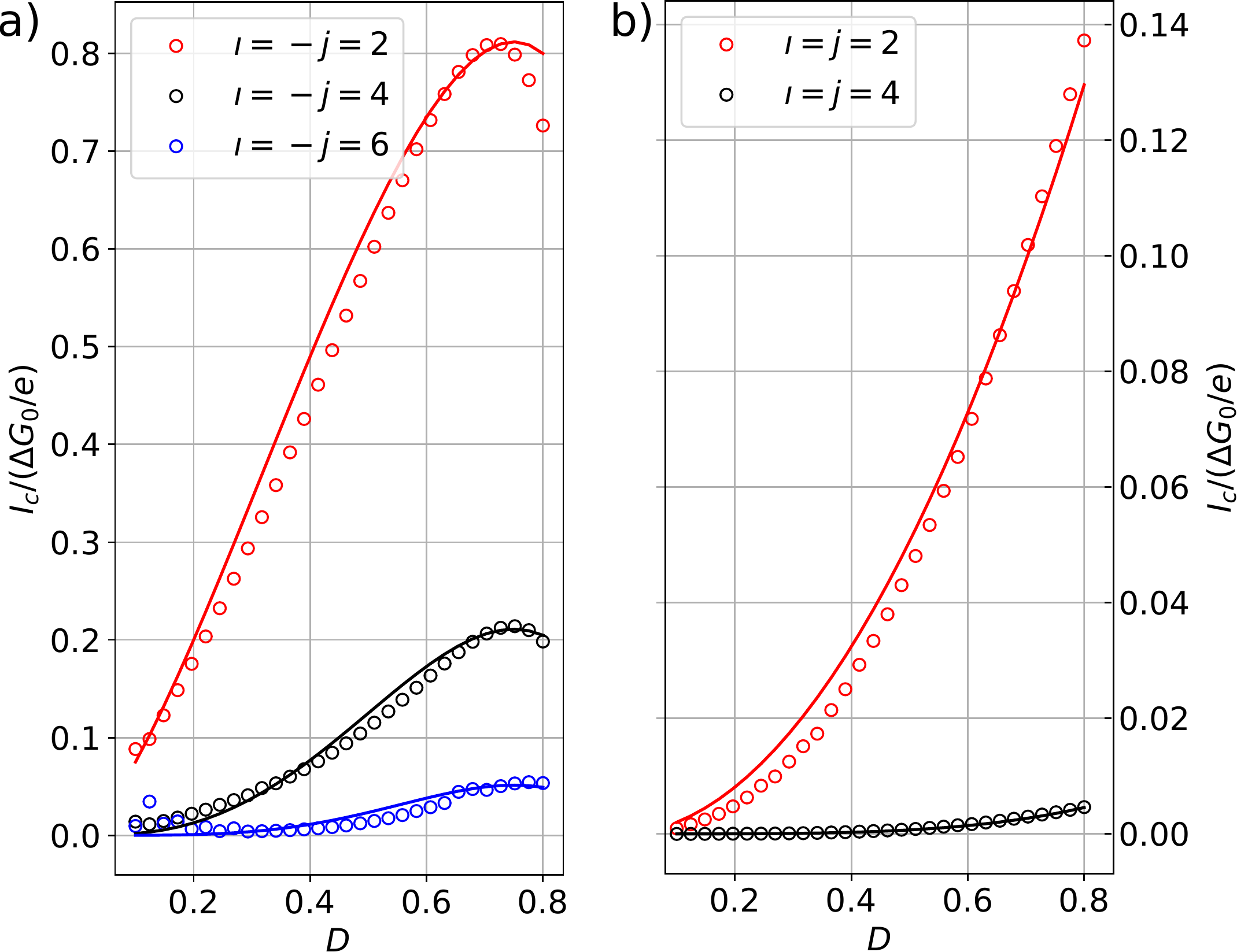}
\caption{Critical current between second and third lead versus the junction transparency D for $eV_2=eV_3=1.4\Delta$ (a) and $eV_2 = -eV_3=1.4\Delta$ (b). The open circles present the numerical results and the curves are analytical dependencies as described in the text.}
\label{IcvsD}
\end{figure}

\begin{figure*}[ht!]
\center
\includegraphics[width = 18cm]{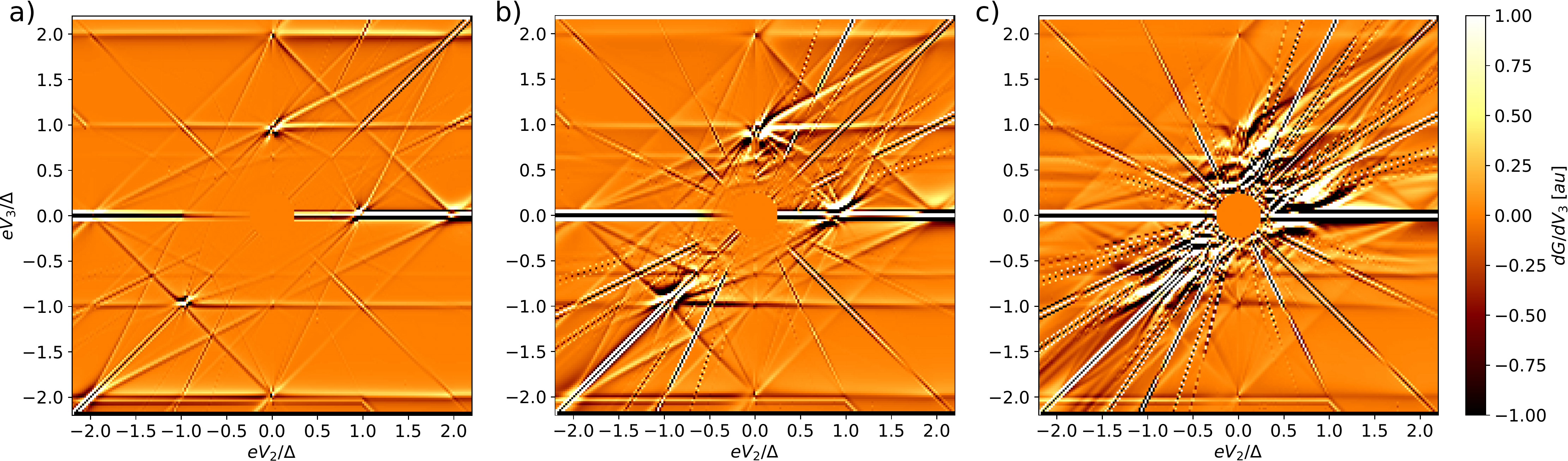}
\caption{Second derivative of the current in the first lead for $D = 0.25$ (a) $D = 0.5$ (b) and $D=0.75$ (c). The centrifugal lines are enhanced by the supercurrent.}
\label{second_derivative_D025D075}
\end{figure*}

To elucidate the supercurrent dependence on the junction transparency we now turn our attention to the dependence of the maximum of the supercurrent components on $D$. In Fig. \ref{IcvsD} (a) with the open circles we show the critical current $I_c = \mathrm{max}[I(\phi_{II})-I(\phi_{III})]$ for $V_2=V_3$, while the solid curves are estimates of the critical current inferred from the junction transparency. For $\imath=-j=2$ we have $I_c \simeq a_1 \sqrt{D^2}\cdot \sqrt{D(1-D)}$ according to the analysis of the amplitudes of the wave-functions in the second lead. For $\imath =-j=4$ we have $I_c \simeq a_2 \sqrt{D^2\cdot D(1-D)}^2$ and finally for $\imath=-j=6$ $I_c \simeq a_3 \sqrt{D^2\cdot D(1-D)}^3$ with the free parameters $a_i$ chosen for the best fit. We see that the higher order components of the current require more scattering events between the biased and unbiased electrodes, so that the wave-functions acquire multiple gains in phase---hence the decreased oscillation period in current-phase relation. At the same time the higher order components are more sensitive to the transparency of the normal part---resulting in the increased power of $I_c$ dependence on $D$. 

We also calculate the critical current for $V_2=-V_3$ as a function of $D$ and find that it follows the same rule of increasing power in $D$ for increasing component number [see Fig. \ref{IcvsD} (b)]. It is important to note here that the detailed shape of the curve---such as the presence of a local maximum---depends on the voltage for which we obtain the curve. Nevertheless the qualitative dependence on the power of $D$ is independent on that choice. 

Finally, let us go back to the analysis of junction response when both bias voltages are varied. To highlight the subharmonic features we now present the second derivative of the current in the first lead versus the bias voltages for low transparency [Fig. \ref{second_derivative_D025D075} (a)], average transparency [Fig. \ref{second_derivative_D025D075} (b)] and high transparency [Fig. \ref{second_derivative_D025D075} (c). Comparing the maps we observe a distinct change in the visibility of both: the non-local MAR lines and centrifugal supercurrent-enhanced lines. Making the system more transparent results in smoothing out the non-local MAR features and significant amplification of the visibility of the current consisting of multiple DC components. The latter results from the presence of higher order DC components that require many sequential scattering events through the junction and are favored in transparent junctions. On the other hand the usual MAR contribution to the map lacks higher order components and hence resembles the usual MAR response to the transparency---the lower the transparency the sharper the features.

\section{Discussion}
The conductance map obtained within this work shows a similar subharmonic gap pattern as the map Fig. 1 of Ref. \onlinecite{houzet_multiple_2010}. Here however, we are able to recover the full spectrum of the features---in particular with those at commensurate voltages---missing in the modeling of Ref. \onlinecite{houzet_multiple_2010} due to incoherent nature of the transport. On the other hand the obtained supercurrent features are compatible with the voltage induced Shapiro steps that were demonstrated for a diffusive Josephson junction with a superconducting tunnel probe [\onlinecite{cuevas_voltage-induced_2007}]. By that we are able to recover the full spectrum of features as found in the experiment on the nanowire Josephson junction.

The maps of the second derivative of the current Fig. \ref{second_derivative_D025D075} obtained within the presented model exhibit the features compatible with those in the experimental maps of Ref. \onlinecite{cohen_non-local_2016} [see Fig. 4 and S6 therein]. Specifically, we obtain subharmonic MAR structure but most importantly also the rapid amplification of the current for the commensurate voltages which in the experimental paper was attributed to the supercurrent driven by Andreev bound states in the junction. Here, however we see that all the states in the junction have a finite life-time due to unequal bias voltages. The current amplification appear due the higher order DC current components in which the Cooper pairs are transported between commensurately biased leads by non-equilibrium electrons and holes through MAR processes. When the bias voltages are detuned from the commensurate condition $\imath V_2+jV_3 \ne 0$ the higher order components become rapidly oscillating and drop out of the sum for the current Eq. (\ref{total_current_formula}) resulting in a change in the DC current.

\section{Summary}
We studied coherent multiple Andreev reflections in a multiterminal Josephson junction with two bias voltages. We analyzed maps of the current running through the junction and identified pronounced lines of enhanced current compatible with the ones measured in the recent experiment [\onlinecite{cohen_non-local_2016, cohen_nonlocal_2018}]. This enhancement is the result of multiple DC components contributing to the current for commensurate bias voltages. 
We found that the principal DC component of the current is voltage-dependent and corresponds to the dissipative current, while the higher order ones are non-dissipative and depend on the time-independent phase applied to the superconducting leads. This happens despite the absence of bound states that may carry dissipationless current. We explained that the latter components result from nonlocal Andreev reflections of non-equlibrium particles propagating through the junction. We found that the DC current component order $n$ determines both: the periodicity of the current/phase relation with the period $4\pi/n$ and the magnitude of the supercurrent with the main trend given by $\sim D^n$. Finally, we identified the transparency dependence of the enhanced current and MAR lines in current maps and point out the open systems as the preferable one for observation of the supercurrent-enhanced lines.

\section*{Acknowledgments}
We are thankful to D. C. Sticlet, M. Irfan and T. {\" O}. Rosdahl for helpful discussions. This work was supported by National Science Centre, Poland (NCN) according to decision number DEC-2016/23/D/ST3/00394, by the Netherlands Organization for Scientific Research (NWO/OCW) and by the European Research Council (ERC). The calculations were performed on PL-Grid Infrastructure. 

\bibliography{mmar}

\end{document}